\documentclass[]{aa}
\usepackage{graphicx}
\usepackage{txfonts}
\usepackage{amsmath}
\usepackage{comment}

\newcommand{\Omegab}{\Omega_\mathrm{b}}

\newcommand{\Kpc}{~\mathrm{kpc}}

\newcommand{\kmsec}{~\mathrm{km}~\mathrm{s}^{-1}}

\newcommand{\kmseckpc}{~\mathrm{km}~\mathrm{s}^{-1}~\mathrm{kpc}^{-1}}

\newcommand{\vc}{v_{\mathrm{c}}}

\newcommand{\RNum}[1]{\uppercase\expandafter{\romannumeral #1\relax}}

\newcommand{\degree}{^{\circ}}

\newcommand{\pare}[1]{\left(#1\right)}

\newcommand{\Js}{J_{\mathrm{s}}}

\newcommand{\Jf}{J_{\mathrm{f}}}
\newcommand{\Jp}{J_{\mathrm{p}}}

\newcommand{\ths}{\theta_{\mathrm{s}}}
\newcommand{\thf}{\theta_{\mathrm{f}}}
\newcommand{\thp}{\theta_{\mathrm{p}}}

\begin{document}

   \title{Tracing Hercules in Galactic azimuth  
   with Gaia DR2}

   \author{
    G. Monari \inst{1,2}
    \and
    B. Famaey \inst{2}
    \and
    A. Siebert \inst{2}
    \and 
    O. Bienaym\'e \inst{2}
    \and
    R. Ibata \inst{2}
    \and 
    C. Wegg \inst{3}
    \and
    O. Gerhard \inst{4}
   }

    \institute{Leibniz Institut fuer Astrophysik Potsdam (AIP), An der Sterwarte 16, 14482 Potsdam, Germany \\
    \email{gmonari@aip.de}
    \and
    Universit\'e de Strasbourg, CNRS UMR 7550, Observatoire astronomique de Strasbourg, 11 rue de l'Universit\'e, 67000 Strasbourg, France
    \and
    Laboratoire Lagrange, Universit\'e C\^ote d'Azur, Observatoire de la C\^ote d’ Azur, CNRS, Bd de l’Observatoire, 06304 Nice, France
    \and
    Max-Planck-Institut f\"{u}r extraterrestrische Physik, Gießenbachstraße 1, 85748 Garching bei M\"{u}nchen, Germany}

   \date{Received xxxx; accepted xxxx}

% \abstract{}{}{}{}{} 
% 5 {} token are mandatory
 
  \abstract
   { The second data release of the Gaia mission has revealed, in stellar velocity and action space, multiple ridges, the exact origin of which is still debated. Recently, we demonstrated that a large Galactic bar with pattern speed 39 km/s/kpc does create most of the observed ridges. Among those ridges, the Hercules moving group would then be associated to orbits trapped at the co-rotation resonance of the bar. Here we show that a distinctive prediction of such a model is that the angular momentum of Hercules at the Sun's radius must significantly decrease with increasing Galactocentric azimuth, i.e. when getting closer to the major axis of the bar. We show that such a dependence of the angular momentum of trapped orbits on the 
   azimuth would on the other hand not happen close to the outer Lindblad resonance of a faster bar, unless the orbital distribution is still far from phase-mixed, namely for a bar perturbation younger than $\sim$2 Gyr. Using Gaia DR2 and Bayesian distances from the StarHorse code, and tracing the average Galactocentric radial velocity as a function of angular momentum and azimuth, we show that the Hercules angular momentum changes significantly with azimuth as expected for the co-rotation resonance of a dynamically old large bar.
   }

   \maketitle
%
%________________________________________________________________

\section{Introduction}

It has long been known that the distribution of stars in local velocity space in the solar vicinity is very far from a smooth velocity ellipsoid \citep[e.g.,][]{Chereul1998,Dehnen1998,Famaey2005}. This observational fact has become more striking than ever with the data from the second data release of the Gaia mission \citep{Gaia}, revealing in impressive detail prominent ridges in local velocity and action space. One of these ridges is associated to the long known structure dubbed the Hercules moving group, whose exact origin remains debated. For 20 years, this moving group has been suspected to be associated to the resonant interactions of local stars with the central Galactic bar \citep{Dehnen1999bar,Dehnen2000}. For instance, if the Sun is located just outside the outer Lindblad resonance (OLR) of the bar, where stars make two epicyclic oscillations while making one retrograde rotation in the rotating bar frame, a Hercules-like moving group is generated by the linear deformation of the axisymmetric background phase-space distribution function \citep{Monari2017a, Monari2017}. This implies a bar pattern speed around $\Omegab \simeq 55~\kmseckpc$ \citep[see also, e.g.,][]{Fux2001,Chakrabarty2007,Minchev2007,Minchev2010,Quillen2011,Antoja2014,Fragkoudi2019}. However, \citet{PerezVillegas2017} demonstrated that orbits trapped at the co-rotation resonance of the bar could also reproduce the Hercules moving group in local velocity space. This explanation would be more in line with the pattern speed deduced from recent dynamical modelling of the stellar \citep[][]{Portail2017, Sanders2019, Clarke2019} and gas kinematics \citep{Sormani2015,Li2016} in the bar/bulge region, yielding $\Omegab \sim 37-41~\kmseckpc$. In particular, \cite{Monari2019} (hereafter M19) have shown that the Galactic model of \cite{Portail2017} can reproduce -- from the resonances with the bar alone  -- most of the observed features of the local velocity and action space: while Hercules is related to the co-rotation resonance, the so-called `horn' feature \citep{MonariPhD,Monari2017a,Fragkoudi2019} is then associated to the 6:1 resonance of the $m=6$ mode of the bar potential, and the high-velocity arch or `hat' \citep[e.g.,][]{Hunt2018} is caused by the OLR of the $m=2$ mode.

However, most of these structures in the phase-space distribution of the solar neighbourhood can be explained by different combinations of non-axisymmetric perturbations, making their modeling degenerate \citep{Hunt2019}. For instance, if the Sun is located just outside of the bar's OLR, Hercules is caused by the linear deformation of velocity space and orbits trapped at the OLR are responsible for the `horn' feature which delineates the boundary of the linear deformation zone \citep[e.g.][]{Fragkoudi2019}, whilst if the Sun is located just outside the bar's co-rotation radius, orbits {\it trapped} at co-rotation are responsible for Hercules and the `horn' corresponds to stars trapped at the 6:1 resonance (M19). This kind of degeneracy persists even if we extend our radial coverage: for instance, looking at the ridges in tangential velocity space vs. Galactocentric radius \citep{Monari2017,Ramos2018, Laporte2019,Fragkoudi2019}, or in action space \citep{Trick2019a}, it is not clear how to discriminate between different models \citep{Trick2019b}. However, following the ridges as a function of azimuth should in principle be a promising way to disentangle the effect of different resonances. \cite{FriskeSchonrich2019} for instance analyzed the ridges in the average Galactocentric radial velocity as a function of angular momentum and azimuth, not limiting themselves to the Sun's Galactocentric radius. They concluded that all ridges taken together, if they would belong to a single pattern, would correspond to an azimuthal wavenumber $m=4$. 

In this Letter, we concentrate on stars in an annulus of 400~pc width, centered around the Sun, and we follow the non-zero radial velocity ridge corresponding to Hercules in angular momentum vs. azimuth at this Galactocentric radius, as a potential discriminant between different models for the origin of the Hercules moving group. In Sect.~2, we describe our modelling procedure for computing the stellar phase-space distribution function in the zones of resonant trapping, and the expectations for the slope of the ridge in angular momentum vs. azimuth for the OLR and co-rotation resonances respectively. In Sect.~3, we compare these theoretical expectations with Gaia data, and we conclude in Sect.~4.

\section{Analytical modelling}\label{sec:model}

\subsection{Reduction to a pendulum}

To describe the stellar phase-space distribution function (DF) in the zones of resonant trapping, we use the perturbation theory method presented in \citet[][M19]{Monari2017b}. Hereafter, the potential of the Galaxy is the one described in \cite{Portail2017}, used precisely as described in M19.

The natural phase-space coordinates to study stellar dynamics in the Galaxy are action-angle variables.  In an axisymmetric Galaxy within a cylindrical coordinate system $(R,\phi,z)$, the three natural action coordinates to use are $J_R, J_\phi, J_z$ and the canonically conjugate angles evolve linearly with time as $\theta_i(t) = \theta_{i0} + \omega_i t$, where the fundamental orbital frequencies $\omega_i$ are functions of the actions. In an equilibrium configuration, the angle coordinates of stars are phase-mixed on orbital tori that are defined by the actions alone, and the unperturbed distribution function $f_0$ is a function of actions alone (Jeans theorem). The angles, on the other hand, indicate where each star is along its orbit, and in particular the angle $\theta_\phi$ is closely related to the azimuth $\phi$ of the star. Hereafter, we work with the epicyclic approximation to estimate actions and angles \citep[e.g.][]{Monari2016}.

We then consider a perturbation by a Galactic bar rotating at pattern speed $\Omegab$. At a resonance with the bar, the fundamental orbital frequencies $\omega_R$ and $\omega_\phi$ are such that
\begin{equation}
  l\omega_R+m(\omega_\phi-\Omegab)=0.
\end{equation}
The co-rotation resonance corresponds to $l=0$. For the $m=2$ mode of the bar, the OLR corresponds to $l=1$. 

The key canonical transformation in our procedure is the one going from the axisymmetric angle and action variables to the ``slow'' and ``fast" variables
\begin{equation}\label{eq:canonical_transf}
\begin{aligned}
  \ths & =l\theta_R+m\pare{\theta_\phi-\Omega_b t},\quad & \Js& = J_\phi/m, \\
  \thf &= \theta_R, \quad & \Jf &= J_R - (l/m) J_\phi.
\end{aligned}
\end{equation}
As is evident from the definition of the frequencies and that of the resonance, the ``slow'' angle $\ths$ is said to be slow because it evolves very slowly close to the resonance. For the co-rotation resonance, $\ths$ is very nearly the azimuthal angle in the bar frame. Near the resonance the ``slow" action (namely a fraction of the specific angular momentum $J_\phi$) also evolves slowly, and one can average the Hamiltonian over the fast angles. Formally, such an averaging produces an exactly constant fast action. For each such fast action, the Hamiltonian for the motion of the slow angle then becomes that of a pendulum, with the slow action as canonically conjugated momentum. One can define the energy of that pendulum and thus determine whether it is ``librating" or ``circulating". When the pendulum is librating, the associated orbit is said to be ``trapped at the resonance''. For such a librating pendulum, one then makes a new canonical transformation defining the actual action-angle variables of the pendulum itself. The trapped DF close to the resonance can then be defined as the original DF phase-mixed over those pendulum angles.

\subsection{Azimuthal variations of different resonant zones}

Let us now heuristically determine -- for different resonances with the bar's $m=2$
mode -- how the location of the trapping zone moves in local velocity space when changing the azimuthal angle to the bar in configuration space. For this, let us consider a set of trapped orbits at the azimuth of the Sun within a small annulus of Galactocentric radii around the Sun, and hence with different radial angles $\theta_R$, but with similar angular momentum $J_\phi$. Let us now evolve these orbits in time. The radial angle is a fast-varying variable ($\theta_R=\theta_f$), hence once the radial angles have varied by a small amount, e.g. $\sim \pi/4$, along any of these orbits, their slow angle $\ths$ and action $J_s$ will have varied only very little.

If we consider the $(l,m)=(1,2)$ OLR, Eq.~2 then implies that the azimuthal angle in the bar frame will have evolved in magnitude by $\sim \pi/8 = 22.5 \degree$, whith almost constant $\ths$ and $J_s$. Hence we expect that orbits trapped at the OLR will remain at almost constant $J_\phi=2 \Js$ over that range of  azimuthal angles to the bar in the observed data. Note that this heuristic argument becomes less and less valid for resonances with much larger values of $m$. 

On the other hand, if we consider the $(l,m)=(0,2)$ co-rotation resonance, the angle to the bar becomes half the slow angle because $l=0$, and this means that any significant variation of this angle to the bar will be accompanied by a similar variation (and even larger by a factor of 4) in the angular momentum $J_\phi$. Hence we expect the angular momentum of orbits trapped at co-rotation to significantly vary with azimuth, while this is not the case for stars trapped at the OLR. 

In summary, at the OLR, the azimuthal angle of stars in the bar frame varies more quickly than their angular momentum, whilst at co-rotation the azimuth of trapped orbits in the bar frame varies very slowly, and such variations have to be accompanied by a significant change of angular momentum.

\subsection{Computing the trapped distribution function}

\begin{figure}
    \centering
    \includegraphics[width=\columnwidth]{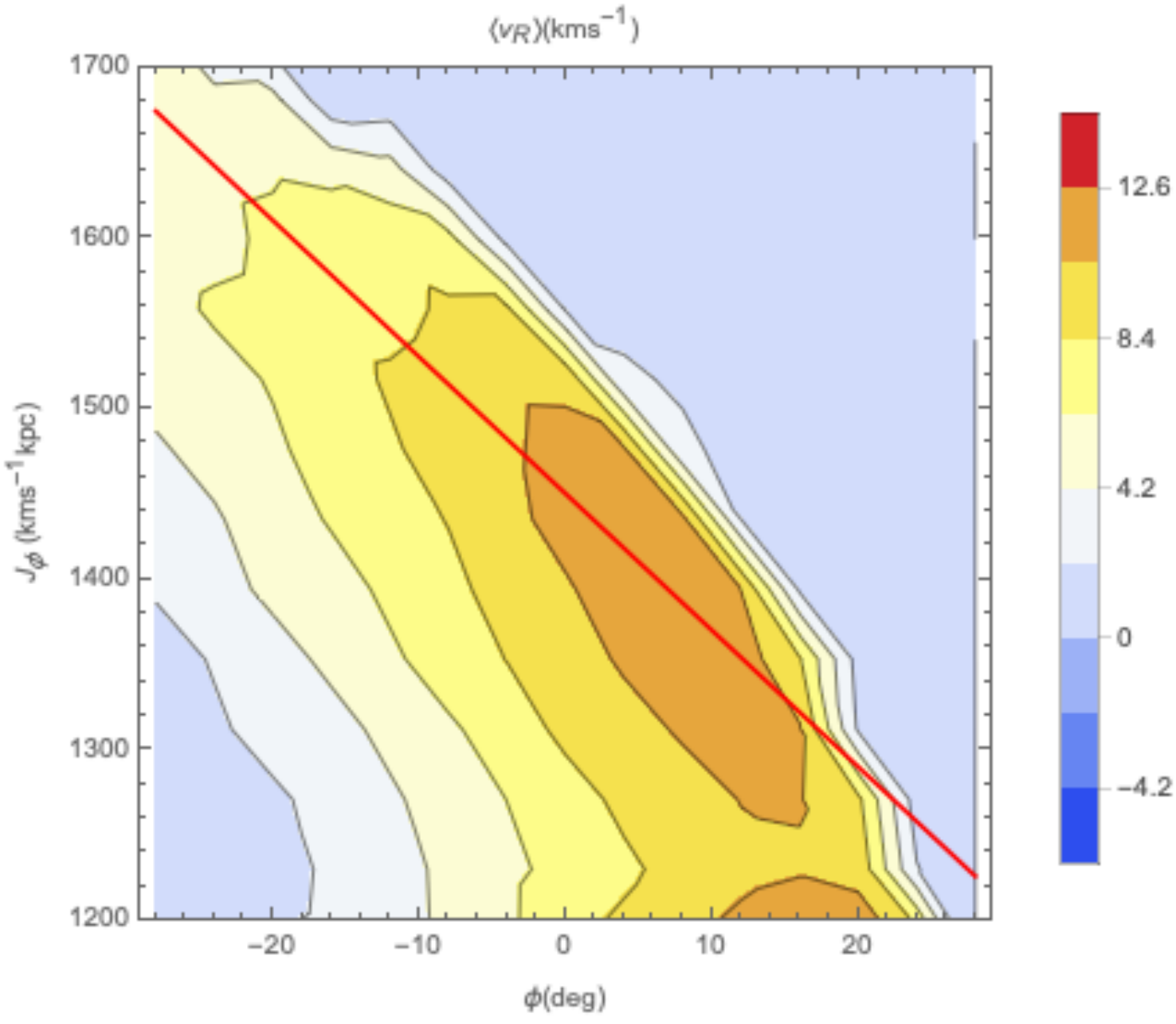}
    \caption{Mean $v_R$ in the $(\phi,J_\phi)$ space obtained from DFs computed on velocity grids ($\Delta v =10~\kmsec$) at $R=R_0$ and different $\phi$ at a distance $\Delta \phi = 4\degree$, using the \cite{Portail2017} model and the method described in \cite{Monari2017b} and M19 (see Sect.~2 for details). The red line corresponds to a slope of $-8~\kmsec\text{deg}^{-1}$.}
    \label{fig:phiJphi39th}
\end{figure}

Let us now verify the above heuristic argument of Sect.~2.2 by making a quantitative evaluation of the trapped DF using the same method as in M19. First, we define an unperturbed DF $f_0$ that is a reasonable representation of the background distribution of disk stars in phase-space (i.e. neglecting deviations from axisymmetry, moving groups, etc.). We take it to be the quasi-isothermal DF $f_0(J_R,J_\phi)$ \citep{Binney2010} used in M19 (see M19 for details). Then we use the action-angle variables $(\Js,\Jf,\ths,\thf)$, defined in Eq.~\eqref{eq:canonical_transf}, to determine the regions of phase-space where orbits are trapped to resonances (librating pendula) and where they are not (circulating pendula). As stated before, in the vicinity of the resonant regions, the dynamics of the $(\ths,\Js)$ variables is that of a pendulum, for which we can also define action-angle variables $(\thp,\Jp)$. We derive the DF for the trapped/librating and circulating stars in the same way as explained in \cite{Monari2017b} and M19: 1) for the librating stars we take $f = \langle f_0(\Jf,\Js(\thp,\Jp)) \rangle$ , 2) for the circulating stars  $f = f_0(\Jf,\langle \Js(\thp,\Jp) \rangle)$, where $\langle \cdot \rangle$ represents the average along $\thp$ from $0$ to $2\pi$. This allows us to compute $f$ at different positions and velocities, directly related to actions and angles using the epicyclic approximation \citep[see, e.g.,][]{Monari2016}. We select points in configuration space at $R=R_0$ \citep[$R_0=8.2~\Kpc$ in the][model]{Portail2017}, with different azimuths $\phi$. The azimuth is $\phi=0$ at the Sun's position and positive in the direction of Galactic rotation, such that a point at $\phi=28\degree$ is aligned with the long axis of the bar in the model. In particular, we consider points equispaced in $\phi$ and at a distance $\Delta\phi = 4\degree$. For each of these points we compute $f$ on a grid in velocity space of bin-size $\Delta v = 5 \kmsec$, for $-150~\kmsec< v_R < 150~\kmsec$, and $80~\kmsec < v_\phi < 320~\kmsec$. We then use these grids to obtain the mean $v_R$ in the $(\phi,J_\phi)$ space: at each of the $\phi$ points considered, $J_\phi$ is given by $J_\phi = R_0 v_\phi$, while the mean $v_R$ is obtained summing all the values of $v_R$ on the velocity grid corresponding to a particular $v_\phi$  (or $J_\phi$), weighted by $f/f_0$. In this way, and interpolating on the grid in $(\phi,J_\phi)$, we obtain Fig.~\ref{fig:phiJphi39th}. 

Fig.~\ref{fig:phiJphi39th} confirms in a rigorous way what we explained hereabove with the heuristic argument. In this space, a clear ridge of positive $v_R$ appears, associated to the Hercules moving group in the model (see M19). In this model, Hercules is formed by stars trapped to the co-rotation, which, according to our heuristic argument, vary significantly in $J_\phi$ as one varies the $\phi$ angle. In particular, the ridge is inclined such that Hercules shifts to lower $J_\phi$ as we move in $\phi$ towards alignment with the long axis of the bar. We note that it also becomes less populated, as recently shown in the $N$-body simulations by \citet{Donghia2019}. These self-consistent simulations actually also displayed a displacement of Hercules with azimuth at co-rotation, which our present heuristic argument and analytic model allows us to fully confirm and physically explain. To quantify the shift in $J_\phi$ of the co-rotation angular momentum ridge with azimuth, we overplot to the ridge a line of slope $-8~\kmsec\Kpc~\text{deg}^{-1}$ to give a rough quantitative (and yet simple) description of this trend with $\phi$. Note that only the slope is relevant for the comparison we intend to make with the data, as the zero-point of the relation depends on details such as the peculiar velocity of the Sun and the slope of the Galaxy rotation curve (see M19 for details), whilst the exact value of the average $v_R$ depends on the choice of estimate of the action-angles (we used the epicyclic approximation here) and the form of the background DF $f_0$.

To complete the check of the heuristic argument, we also studied the behaviour of the model when increasing the pattern speed parameter to $\Omegab = 50~\kmseckpc$, so that the OLR is slightly inside the Sun's circle\footnote{To be completely consistent, one should also vary the length of the bar, so that it does not extend outside the CR circle. However, this is complicated to do in a model like the one of \cite{Portail2017} and the main point regarding the azimuthal behaviour of the Hercules moving group does still hold even if the model is not completely consistent.}. Also in this case ridges of negative and positive mean $v_R$ are formed in the proximity of the OLR position, and correspond in this model to the Hercules moving group (formed by stars {\it outside} the trapping region and linearly deformed by the OLR) and the horn (formed by stars {\it inside} the trapping region). In both cases the variation of the position of the ridges with $\phi$ is so insignificant that the ridges appear as almost straight lines of constant $J_\phi$, with zero slope.

\section{Observational data from Gaia DR2}

We now consider stars from the subset of Gaia DR2 for which RVS l.o.s. velocities are available ($\sim 7$ million stars). We use the photo-astrometric distances to these stars provided by \cite{Anders2019}, obtained using the code \verb!StarHorse!. We consider only the subset of stars in the sample that respects the flags recommended by the authors which corresponds to the most robust distance estimates, i.e., \verb!SH_GAIAFLAG=''000''! and \verb!SH_OUTFLAG=''00000''!. After the cleaning we are left with 6350087 stars. 

\begin{figure}
    \centering
    \includegraphics[width=\columnwidth]{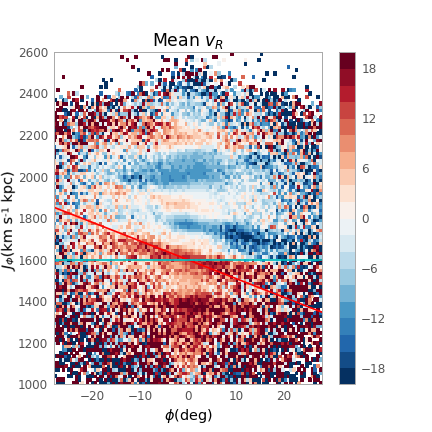}
    \caption{Mean $v_R$ in the $(\phi,J_\phi)$ space obtained for stars from Gaia DR2, with distances estimated with StarHorse, inside an annulus of size $\Delta R = 0.4~\Kpc$ around $R=R_0$. The bin sizes are $\Delta \phi = 0.56\degree$ and $\Delta J_\phi = 16~\kmsec\Kpc $. The red and blue lines correspond respectively to slopes of -8~$\kmsec\Kpc~\text{deg}^{-1}$ (expected for CR) and $0~\kmsec\Kpc~\text{deg}^{-1}$ (expected for the OLR).}
    \label{fig:phiJphiData}
\end{figure}

We derive for these stars Galactocentric positions and velocities in cylindrical coordinate $(R,\phi,z)$, and $(v_R,v_\phi,v_z)$, %$=(\dot{R},R\dot{\phi},\dot{z})$,  
assuming the parameters for the Sun's distance from the Galactic centre $R_0 = 8.2~\Kpc$, very close to the value of \citet{Gravity2019}, Sun's peculiar velocity w.r.t. the Local Standard of Rest $(U_\odot,V_\odot,W_\odot) = (11.1,12.24,7.25)~\kmsec$, and Galaxy's circular velocity at the Sun $\vc = 233.1~\kmsec$ \citep{Schonrich2010,McMillan2017}. We select stars in an annulus of $R$ such that $|R-R_0|<0.2~\Kpc$, leaving us with 1535484 stars, and we plot the mean $v_R$ of these stars as a function of $\phi$ (same convention than in Section~\ref{sec:model}) and $J_\phi$, calculating mean $v_R$ in bins of size $\Delta \phi = 0.56\degree$ and $\Delta J_\phi = 16~\kmsec\Kpc $. We show this in Fig.~\ref{fig:phiJphiData}.

The data show prominent ridges, similar to those showed by \cite{FriskeSchonrich2019} and \cite{Trick2019b}, but this time only for stars in the small annulus. In particular the ridge corresponding to the Hercules moving group shows a displacement in $J_\phi$ with $\phi$ similar to the one that we showed in the model of Section~\ref{sec:model}. We overplot on the top of it the slopes obtained from the models with trapping at CR (red line) and OLR (blue line), which show a clear agreement with the former case.

This thus demonstrates that the slope of the variation of the angular momentum of Hercules with azimuth is qualitatively in line with the signature of the co-rotation resonance of a dynamically {\it old} large bar like that of \citet{Portail2017} and \citet{Clarke2019}. A key assumption of our analytical modelling is indeed that trapped orbits are phase-mixed along the pendulum angles. In the case of an OLR origin of Hercules, the only way out would thus be to drop this assumption, hence considering that the bar is still fairly young, such that orbits are still far from phase-mixed in the bar frame. As is well known \citep[e.g.,][]{Minchev2010}, due to the initial response of the disk, a bar younger than $\sim$2~Gyr can indeed cause transient features in local velocity space and displace its OLR signature w.r.t. the equilibrium phase-mixed case. This could indeed cause a significant variation of the angular momentum with azimuth even at the OLR, but only within the first $\sim$2~Gyr after bar formation \citep[see, e.g., Fig.~7 of][]{Trick2019b}.

\section{Discussion and conclusion}

In this Letter, we analytically demonstrated that the resonant zone associated to the co-rotation resonance of the Galactic bar should significantly vary in angular momentum as a function of azimuth at a given Galactocentric radius, whilst this should not happen for the OLR. Using Gaia DR2 and Bayesian distances from the \verb!StarHorse! code \citep{Anders2019}, and tracing the average Galactocentric radial velocity as a function of angular momentum and azimuth, we then showed that the angular momentum of the Hercules moving group changes significantly in azimuths ranging from $-20\degree$ to $20\degree$ (in an annulus of 400~pc width around the Galactocentric radius of the Sun), as expected for orbits trapped at co-rotation in the model of M19, based on the large bar of \citet{Portail2017}. 

Our findings thus reinforce the case for a co-rotation origin of the Hercules moving group in the case of a dynamically old bar ($>2$~Gyr). The only way for an OLR origin of the Hercules moving group to explain the observational trend reported here would be if orbits are still far from phase-mixed in the bar potential, namely for a bar perturbation younger than $\sim$2 Gyr. But our Fig.~2 also raises a certain number of interesting new questions that should be addressed in subsequent studies. First of all, as is well known since the publication of the Gaia DR2 catalogue \citep[e.g.,][]{Katz2018,Trick2019a}, Hercules has a secondary lower angular momentum component, whose origin is still unclear \citep[e.g.,][]{Li2019}. This lower angular momentum ridge actually appears less inclined than the main Hercules ridge on Fig.~2, and might thus have a different origin. In addition, the `horn' feature, which appears as the ridge right above Hercules on Fig.~2, would correspond to the 6:1 resonance with the $m=6$ mode of the bar in the model of M19. The fact that the radial velocity of the horn appears to vary very quickly with azimuth on Fig.~2, with a strongly negative radial velocity at positive azimuth and a complete disappearance of the ridge at negative azimuth, favors a resonance with such a high order mode. However, it is not clear whether the amplitude of the $m=6$ mode of the bar would be sufficient to create such a prominent feature at positive azimuth. Intriguingly, the horn ridge also appears inclined: while one expects the ridges to become more inclined with higher order resonances than for the 2:1, the fact that the slope does not seem too different from the Hercules slope is intriguing. There thus remain numerous intriguing features of our Fig.~2 to understand in future works, and it is clear that spiral arms and the recent interaction of the disk with the Sagittarius dwarf galaxy might also play a role in interpreting these kinematic features \citep[e.g.,][]{Laporte2019}. Data in a larger range of azimuths with future Gaia data releases will also allow all these potential effects to be tested further.

\begin{acknowledgements}
The authors thank Elena D'Onghia and Adrian Price-Whelan for useful discussions. BF AS and RI acknowledge support from the ANR project ANR-18-CE31-0006. This project has received funding from the European Research Council (ERC) under the European Union’s Horizon 2020 research and innovation programme (grant agreement No. 834148). This work has made use of data from the European Space Agency (ESA) mission {\it Gaia} (\url{https://www.cosmos.esa.int/gaia}), processed by
the {\it Gaia} Data Processing and Analysis Consortium (DPAC,
\url{https://www.cosmos.esa.int/web/gaia/dpac/consortium}). Funding
for the DPAC has been provided by national institutions, in particular
the institutions participating in the {\it Gaia} Multilateral Agreement.
\end{acknowledgements}

%-------------------------------------------------------------------

\bibliographystyle{aa}
\bibliography{azimuthbib}

\end{document}